# Ascorbic acid mitigates oxidative structural degradation in bovine spermatozoa: a label-free quantitative phase microscopy study

Harpreet Kaur [a,b], Shubham Tiwari [a,b], Aman Kumar [c], Sandeep Kumar Jha[d], Satish Kumar Dubey [a], Dalip Singh Mehta [a,b,#]

[a]Center for Sensors, Instrumentation and Cyber Physical System Engineering (SeNSE), Indian Institute of Technology Delhi, Hauz Khas, New Delhi-110016, India

[b]Bio-photonics and Green-photonics Laboratory, Department of Physics, Indian Institute of Technology Delhi, Hauz-Khas, New Delhi 110016, India.

[c]Lala Lajpat Rai University of Veterinary and Animal Sciences (LUVAS), Hisar, Haryana-125004

[d]Center for Biomedical Engineering (CBME), Indian Institute of Technology Delhi, Hauz-Khas, New Delhi 110016, India

#mehtads@physics.iitd.ac.in

**Abstract**: Oxidative stress is a key factor in low fertility outcomes during assisted reproduction technology (ART) and contributes to poor sperm quality. Conventional sperm assessment relies on bright-field microscopy, which lacks the quantitative sensitivity to resolve subcellular structural and biophysical changes without exogenous contrast agents, a requirement that can introduce cytotoxic effects and compromise cell viability. This study uses quantitative phase microscopy (QPM) as a label-free, high-throughput method to demonstrate how ascorbic acid (Vitamin C) reduces the impact of oxidative structural degradation in Sahiwal bovine spermatozoa. To illustrate this pathology, severe oxidative stress was experimentally induced using hydrogen peroxide ($H_2O_2$), and two doses of ascorbic acid (1 mg/ml and 8 mg/ml) were tested to evaluate dose-dependent antioxidant recovery in structurally damaged sperm. QPM was used to quantify structural changes in key biophysical parameters, including dry mass, optical thickness, volume, surface area, sphericity, and surface area-to-volume ratio, as well as intracellular texture parameters. Under acute oxidative stress, bovine spermatozoa exhibited significant reductions in optical thickness and dry mass, alongside measurable changes in intracellular structural organization, collectively indicative of oxidative stress-induced morphological degradation. Co-treatment with ascorbic acid resulted in partial, dose-dependent structural preservation, with the 8 mg/ml formulation demonstrating statistically significant attenuation of these structural changes compared to the 1 mg/ml treatment group. These findings suggest that QPM-derived biophysical parameters may serve as promising, label-free structural indicators for characterizing oxidative damage in bovine spermatozoa. Future studies should incorporate functional validation to determine whether this structural preservation translates to improved outcomes in ART.

**Introduction**: Infertility is a major global concern in both human reproductive medicine and the livestock industry, with male-factor defects significantly reducing reproductive efficiency [1]. In bovine breeding programs, evaluating sperm quality is essential for determining conception rates and the overall economic success of herd management. To address fertility issues, Assisted Reproductive Technologies (ART), such as artificial insemination (AI), in vitro fertilization (IVF), and intracytoplasmic sperm injection (ICSI), are commonly used [2,3]. Recent estimates suggest that over 2.5 million assisted fertilization cycles are performed annually worldwide; however, average success rates remain approximately 33% [4]. This highlights the limitations of current selection protocols, particularly in techniques such as ICSI [2,5], where selecting a single spermatozoon depends heavily on the operator's ability to assess cellular health with standard optical tools [6]. These methods may accidentally select cells with reduced fertilization potential because they do not assess the negative effects of oxidative stress using light microscopy. Oxidative stress screening is usually not done in IVF labs because the standard test is complex and expensive [7]. Furthermore, implementing rapid diagnostics could replace lengthy, complex, multi-step analytical procedures that require specialized experimental equipment.

Washing sperm from semen, centrifugation, or cryopreservation may alter sperm cell antioxidant capacity, leading to the production of Reactive Oxygen Species (ROS). While low levels of ROS are physiologically necessary to trigger capacitation and the acrosome reaction, high levels can impair sperm function and lead to infertility [8]. Among these ROS, particularly Hydrogen Peroxide ($H_2O_2$) initiates pathological cascades that induce severe oxidative stress, affecting sperm cell integrity and leading to lipid peroxidation, DNA fragmentation, and protein oxidation; these changes collectively compromise membrane fluidity and intracellular integrity [1,9]. Bovine spermatozoa, a critical model for agriculture and andrology, are uniquely vulnerable to oxidative damage due to the high content of polyunsaturated fatty acids (PUFAs) in their plasma membranes and a distinct lack of cytoplasmic antioxidant enzymes [10,11]. These oxidative cascade-driven structural consequences, including loss of intracellular dry mass resulting from protein leakage, reduction in optical thickness due to membrane lipid peroxidation, and alterations in intracellular structural organization, manifest as measurable changes in the optical and biophysical properties of the sperm head, making them ideally suited to quantitative detection by label-free phase-based imaging approaches such as Quantitative Phase Microscopy (QPM).

To address these structural consequences of oxidative damage, the present study was designed with three interconnected aims: to apply QPM as a label-free structural assessment tool for bovine spermatozoa; to evaluate the dose-dependent protective effect of ascorbic acid against $H_2O_2$-induced structural changes; and to establish a reproducible in vitro model of acute oxidative stress and antioxidant recovery. Ascorbic acid was selected as the antioxidant intervention of interest, given its well-documented role as a water-soluble free radical scavenger and electron donor, which constitutes a primary line of defense against ROS in the seminal plasma, where it protects spermatozoa from oxidative membrane and DNA damage [12,13], which constitutes a primary line of defense against ROS in the seminal plasma, where it protects spermatozoa from oxidative membrane and DNA damage [14]. To protect the PUFA of sperm cells from lipid peroxidation, ascorbic acid reproduces lipid-soluble vitamin E and actively neutralizes $H_2O_2$ and other free radicals [12,15]. Based on this, previous studies have shown that adding ascorbic acid to sperm media provides protection, enhances motility, lowers biochemical markers of oxidative damage, and maintains acrosome integrity. Some studies also produced conflicting results. This discrepancy likely stems from a significant methodological gap in how damage and recovery are assessed. So, a definitive study of its effectiveness is needed [12,16–20].

Therefore, a notable gap in the methodology used to assess this damage and the subsequent recovery process. Subjective grading under bright-field optical microscopy should no longer be considered the sole gold standard for sperm evaluation. In the past, semen analysis and antioxidant effectiveness assessments depended on Computer-Aided Sperm Analysis (CASA) [21,22]. While effective for evaluating motility and morphology, these intensity-based techniques are insensitive to the subtle variations in refractive index and intracellular density associated with early-stage oxidative stress. Alternative methods, such as flow cytometry and fluorescent imaging, offer high specificity for detecting DNA damage [23]. However, they require external chemical staining. Since these stains are often cytotoxic or mutagenic, such methods make the analyzed spermatozoa unusable for ART.

To resolve these shortcomings, label-free modalities should be adopted in modern andrology. Therefore, this study introduces Quantitative phase imaging (QPI), an interferometric technique that has become a powerful tool for label-free analysis of sperm, offering high spatial phase sensitivity. QPI measures the optical path delay of light passing through the cell, producing a highly precise subcellular phase map proportional to the optical thickness and intracellular dry mass [24,25]. This method provides a quantitative, label-free assessment of intracellular dry mass and three-dimensional morphological parameters without exogenous contrast agents or fluorescent stains, thereby preserving cell integrity throughout the imaging process. As a non-invasive optical technique, QPM provides structural and biophysical information not accessible by conventional bright-field microscopy, serving as a complementary approach to traditional sperm assessment methods rather than a replacement.

To the best of our knowledge, this is the first study to apply QPM-derived biophysical and texture parameters to quantify the dose-dependent structural protective effects of ascorbic acid supplementation against acute oxidative stress in bovine spermatozoa. The present study aimed to investigate whether QPM-derived biophysical parameters can detect structural changes in bovine spermatozoa subjected to acute oxidative stress induced by $H_2O_2$, and whether ascorbic acid supplementation attenuates these changes in a dose-dependent manner. To achieve this, sperm head morphological reliability was quantitatively assessed using key optical and three-dimensional parameters, including optical path difference, dry mass, volume, surface area, sphericity, and the surface area-to-volume ratio. In addition, intracellular structural complexity was evaluated through a comprehensive set of label-free texture parameters, including mean optical thickness, standard deviation, skewness, kurtosis, energy, and entropy, to enable objective, reproducible characterization of oxidative damage at the subcellular level. Collectively, this approach establishes whether the direct connection of biophysical parameters to oxidative recovery using QPM as a promising, label-free technique for precisely assessing sperm structural integrity and identifying biophysical biomarkers that may support the selection of the most viable spermatozoa for assisted reproductive technology (ART) protocols.

## 2. Materials and Methods

### Phase 1: Sample preparation and treatment

Experiments on spermatozoa were conducted with approval from the institutional animal ethical committee (IAEC protocol number: IAEC/LUVAS/31/07), and all experiments were conducted in accordance with the ARRIVE guidelines. Frozen bovine semen straws were sourced from Lala Lajpat Rai University of Veterinary and Animal Sciences (LUVAS, Hisar, India). The semen straw was carefully thawed in a water bath at 37 °C for 1–2 minutes to maintain viability, then transferred into microtubes for precise handling. Treatment agents were applied immediately following thawing, sample preparation, sample mounting, and imaging. The analysis was performed in phosphate-buffered saline at pH 7.4 and osmolarity 310 mOsm/kg. All microtubes were kept at 4°C between experimental steps to minimize spontaneous structural degradation and were protected from direct light exposure throughout the procedure, particularly for samples containing ascorbic acid. The study confirms the entire experiment within 48 hours of removing the sample from -196ºC storage. The sample was thoughtfully divided into three groups: control, oxidative stress, and antioxidant recovery, to thoroughly evaluate each condition.

### Phase 1(a): Control condition

For the control group, 10 µL of thawed semen was diluted in 760 µL of PBS to yield a final assay volume of 770 µL, resulting in a suitable sperm suspension for QPM-based structural imaging. Figure 1 illustrates the complete experimental workflow, including sample preparation across all treatment groups, optical image acquisition, TIE-based phase map reconstruction, and the morphological and texture parameters extracted for structural assessment.

### Phase 1(b): Exposure to oxidative stress

For the oxidative stress group, 10 µL of sperm suspension was mixed with 10 µL of hydrogen peroxide ($H_2O_2$) (approximately 30% w/v stock solution bought from Thomas Baker, India) and 750 µL of PBS to induce oxidative stress through the production of ROS. Hydrogen peroxide ($H_2O_2$) was selected as an oxidative stress inducer because it produces physiologically relevant ROS that penetrate the sperm plasma membrane, induce lipid peroxidation, impair mitochondrial function, and compromise DNA integrity [26].

### Phase 1(c): Supplementation with Ascorbic Acid

L- Ascorbic acid was bought from Sisco Research Laboratory (SRL), India, with 99.7% purity. To robustly assess the dose-dependent protective effects of ascorbic acid, an acute in vitro oxidative stress and co-incubation model was developed, utilizing a total assay volume of 770 µL. Stock solutions were precisely prepared by dissolving 1 mg and 8 mg of ascorbic acid in 1000 µL of PBS, ensuring consistency and reliability. The ascorbic acid concentrations of 1 mg/ml and 8 mg/ml were selected based on previously published studies evaluating antioxidants within the 1 mg/ml range in sperm models [27]. More specifically, these doses were informed by a bovine semen study that utilized ascorbic acid at concentrations up to 8.5 mg/ml in cryopreservation extenders [28]. The selected concentrations span the lower and upper boundaries of this established range, enabling a robust assessment of dose-dependent structural protection under acute oxidative stress conditions via QPM-derived biophysical parameters. The ascorbic acid concentration was selected based on published studies that reported its antioxidative effect on bovine spermatozoa [28]. The experimental model was adapted from an established $H_2O_2$-based oxidative stress study applied to sperm cells. The final co-incubation mixture was carefully formulated by simultaneously combining 10 µL of bovine sperm suspension, 10 µL of $H_2O_2$, 60 µL of the specific ascorbic acid stock, and 690 µL of PBS. This simultaneous addition was essential for accurately evaluating the antioxidant's immediate extracellular shielding capacity against acute oxidative stress, in which ascorbic acid competes with ROS during the oxidative addition step. The high dose was intentionally selected to impose a stringent oxidative challenge, enabling meaningful differentiation of dose-dependent protective effect across the two different ascorbic acid concentrations. Table 1 summarizes the concentrations used for three treatment groups.

| **Group** | **Sperm** | **PBS** | **$H_2O_2$** | **AA** | **Final Volume** |
|---|---|---|---|---|---|
| Control | 10 µl | 760 µl | — | — | 770 µl |
| $H_2O_2$ only | 10 µl | 750 µl | 10 µl | — | 770 µl |
| $H_2O_2$ + AA (1 mg/ml) | 10 µl | 690 µl | 10 µl | 60 µl (1 mg/ml) | 770 µl |
| $H_2O_2$ + AA (8 mg/ml) | 10 µl | 690 µl | 10 µl | 60 µl (8 mg/ml) | 770 µl |

**Table 1**: Summary of treatment compositions and final reagent concentrations for each experimental group in the bovine sperm oxidative stress model. AA: Ascorbic acid; PBS: Phosphate-buffered saline.

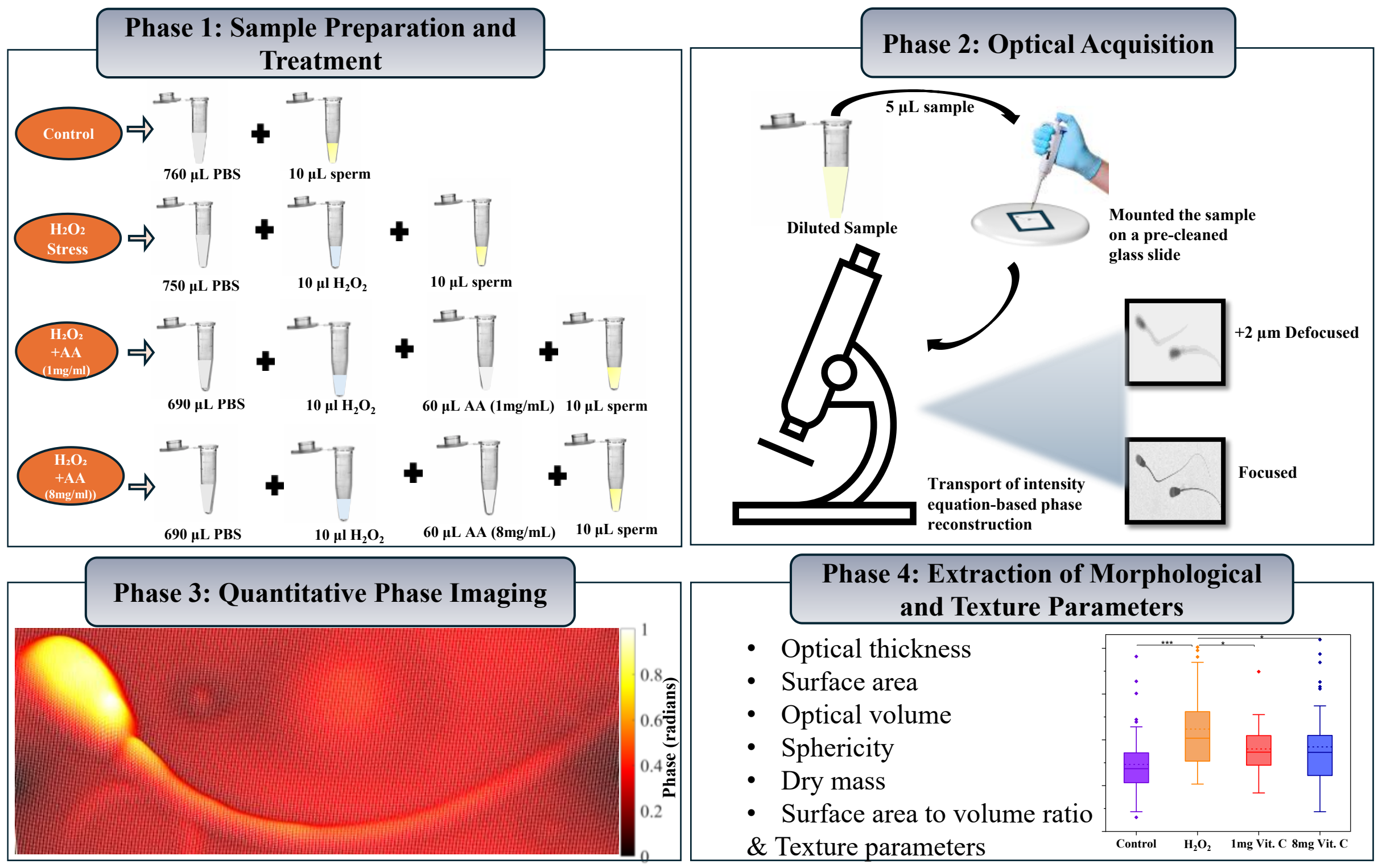

**Figure 1**: Flowchart of the experimental and analytical framework. The methodology is divided into four sequential phases: (Phase 1) Sample preparation comprising four experimental groups Control, $H_2O_2$-induced oxidative stress, and two ascorbic acid co-treatment groups at concentrations of 1 mg/ml and 8 mg/ml respectively; n = 60 cells per group, total n = 240 cells; (Phase 2) Label-free optical profiling performed by placing 5 µL of diluted sample on a pre-cleaned glass slide and acquiring focused and defocused (+2 µm) brightfield images using a bright-field microscope; (Phase 3) Quantitative phase map reconstruction of individual bovine sperm cells using the Transport of Intensity Equation (TIE)-based phase retrieval method implemented in MATLAB; (Phase 4) Quantitative extraction of biophysical metrics, including optical thickness, surface area, optical volume, sphericity, dry mass, and surface area-to-volume ratio and subcellular texture parameters from segmented sperm head regions of interest. Statistical significance between groups was assessed by one-way ANOVA followed by Tukey's post hoc multiple-comparison test ($p < 0.05$, ** $p < 0.01$, *** $p < 0.001$). (AA: Ascorbic acid).

### Phase 2: Optical acquisition

To restrict sperm movement during imaging, a polydimethylsiloxane (PDMS) chamber was placed on a clean glass slide. A 5 µL drop of the sperm suspension was placed inside the chamber and covered with a standard 0.17 mm coverslip. This setup created a flat, uniform layer that optimizes optical transmission and ensures high-quality imaging. The prepared samples were then examined using a bright-field microscope. Finally, to achieve accurate quantitative phase reconstruction, the Transport of Intensity Equation (TIE) method was used, capturing only two images per field of view: one perfectly in focus and one defocused.

### Phase 3: Quantitative phase microscopy

The experimental setup incorporates a Nikon Eclipse LV100 brightfield microscope, a gold standard for sperm cell analysis, ensuring high precision and reliability. Illumination is provided by a broadband tungsten-halogen lamp centered at 540nm, delivering exceptional phase sensitivity during multi-plane intensity acquisition. The system employs Kohler illumination, ensuring uniform light distribution across the field of view and optimal alignment between the field diaphragm and the condenser aperture, thereby enhancing image quality and consistency. A 50×/0.60 NA objective lens captures high-resolution images of individual sperm cells, enabling detailed analysis. Image acquisition is performed with a cutting-edge CMOS camera (Model number C11440-36U, manufactured by Hamamatsu Photonics, with a sensor resolution of 1920 × 1200 pixels and a 5.86 µm pixel size), specifically designed to capture rapid sequences of intensity images without mechanical vibrations, ensuring accuracy and clarity. The experimental setup is shown in Fig. 2. The camera's pixel size ensures perfect Nyquist sampling at 540 nm, matching the objective's optical resolution limits. For phase reconstruction via the TIE, two intensity images are captured along the optical axis: one in focus at z = 0µm and another defocused at z = ±2 µm, and the images are reconstructed using a universal solution-based TIE method [29] as described in detail in Supplementary File 1. This defocus distance is carefully chosen to optimize the finite-difference approximation of the axial intensity derivative, large enough to produce measurable contrast variations that exceed noise levels, yet small enough to maintain the paraxial approximation vital for accurate phase retrieval. Processing these images with a custom TIE-based MATLAB algorithm enables precise reconstruction of the optical phase maps of individual sperm cells. The setup ensures the paraxial approximation remains valid by reconstructing the phase from the magnified image at the detection plane rather than the object plane, minimizing errors from non-paraxial effects and preserving high-frequency spatial details, thereby providing exceptionally accurate phase imaging that advances sperm analysis capabilities.

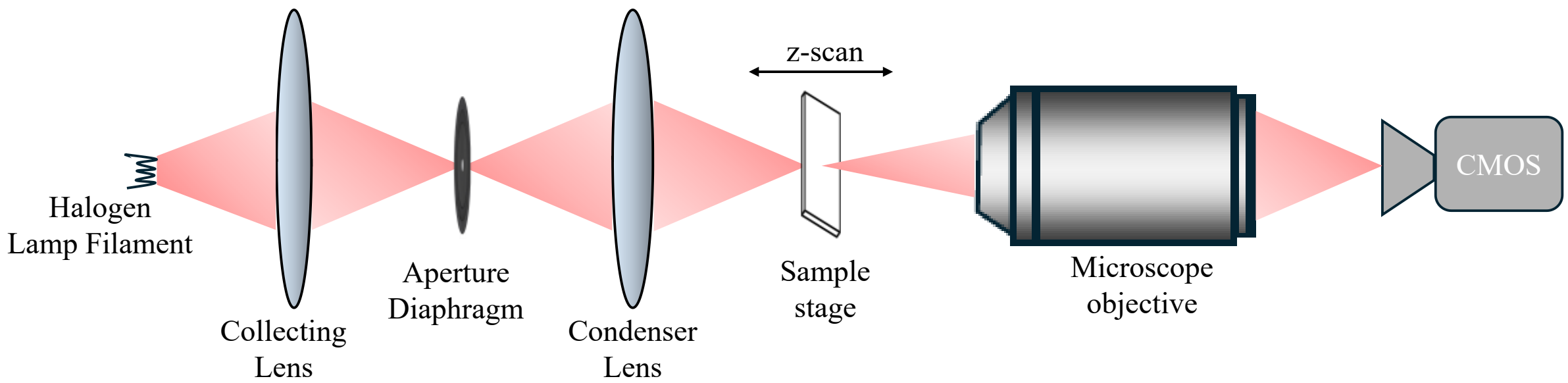

**Figure 2**: Schematic diagram of the bright-field microscope setup used for quantitative phase imaging. Illumination from the light source is directed through a collecting lens, which focuses the light onto the aperture diaphragm. The transmitted light passes through the condenser lens, which directs and concentrates the beam onto the sperm sample mounted on the microscope stage. The motorized sample stage provides ±2 µm axial displacement along the z-axis, facilitating the acquisition of through-focus intensity stacks required for TIE-based phase reconstruction. The transmitted light is subsequently collected by the microscope objective and focused onto the CMOS camera for image capture.

### Phase 4: Extraction of morphological and texture parameters

The quantitative phase map ($\Delta\phi$) generated by TIE-QPM provides a precise 2D projection of the optical path delay caused by spermatozoa. By accurately extracting the sperm head from this map using Tsallis entropy-based thresholding in MATLAB, we can derive critical morphological parameters (optical thickness, dry mass, volume, surface area, sphericity, and surface area-to-volume ratio). and texture parameters (mean, standard deviation, skewness, kurtosis, energy, and entropy) were computed from the optical thickness of the segmented sperm head region of interest, which effectively distinguishes between control, oxidized, and antioxidant-rescued cells, highlighting the method's ability to differentiate and analyze cell states with confidence.

In this study's transmission geometry, the optical thickness, defined as the product of the refractive index difference and the physical thickness, is directly related to the phase shift through the following equation [25,30,31]:

$$Optical\ thickness = \frac{\lambda}{2\pi}(\Delta\phi)$$

where λ represents the wavelength of the illumination source. The optical thickness measurement provides a reliable method for accurately determining the volume of the sperm head through integration over its projected area.

$$Volume = \iint Optical\ thickness\,(x, y)\ dxdy$$

The cellular dry mass represents the total non-aqueous biological components, including proteins and DNA, within the cell. Its precise calculation, using a specific refractive increment value ($\alpha$) of approximately 0.18 mL/g and integrating the phase shift over the projected surface area, provides a reliable measure of cellular biomass. The cellular dry mass is calculated using the following equation:

$$m = \frac{1}{\alpha} \iint \phi(x,y)dx\,dy$$

where α is the specific refractive increment.

The area of the cell surface is calculated based on the gradient of the optical phase delay, utilizing the Monge parameterization method [25,30,31]:

$$S = \iint \sqrt{1 + (\frac{\delta h}{\delta x})^2 + (\frac{\delta h}{\delta y})^2}\, dxdy.$$

Where $\frac{\delta h}{\delta x}$ and $\frac{\delta h}{\delta y}$ are the gradients along the x and y directions of the cell optical thickness profile.

Next, we calculated the sphericity (ψ) of sperm cells, which measures how closely the shape of the sperm head resembles a perfect sphere. It is defined as the surface area to volume ratio, calculated as:

$$\psi = \frac{\pi^{\frac{1}{3}}(6V)^{2/3}}{A}.$$

### 2.6 Statistical analysis

All parameters were carefully extracted and analyzed from the quantitative phase maps using MATLAB. Subsequent statistical analysis and data visualization were performed using OriginPro. To determine the differences among the three experimental groups - Control, Oxidative Stress, and Antioxidant Recovery via Ascorbic Acid, a thorough One-Way ANOVA was performed, followed by Tukey's multiple comparison test to confidently identify specific pairwise differences. All data are presented as Mean ± Standard Deviation (SD). The graphical box plots illustrate the distributions of data, including medians, interquartile ranges, and overall variability, across treatment groups. We considered results statistically significant at $p < 0.05$, with significant values denoted as *$p<0.05$, **$p<0.01$, and ***$p<0.001$.

## 3. Results

### 3.1 Reconstructed phase maps of bovine sperm cells

Cryopreserved semen from a single Sahiwal bull was used throughout the study. QPM was subsequently employed to analyze the biophysical characteristics of 240 cells across four conditions: Control, $H_2O_2$ stressed, and two antioxidant groups (1 mg/ml and 8 mg/ml ascorbic acid). The n = 60 cells per group represent technical replicates at the cellular level, ensuring that observed differences in QPM-derived parameters were attributable specifically to the experimental treatments applied within this controlled model. Representative phase imaging sequences for individual sperm cells from each experimental group are presented in Fig. 3. Figure 3A illustrates a sperm cell under control conditions, where subfigure (a) shows the in-focus intensity image, (b) displays the defocused image acquired at z = +2 µm, and (c) presents the reconstructed quantitative phase map obtained via the TIE algorithm. Only individual, non-overlapping sperm cells with clearly resolvable boundaries were selected for analysis. Cells that contained debris or exhibited imaging artifacts, were excluded to prevent segmentation errors and ensure accurate phase reconstruction. Figure 3B depicts a sperm cell exposed to $H_2O_2$-induced oxidative stress, with panels (d–f) corresponding to the focused, defocused, and reconstructed phase images, respectively. Figure 3C shows a sperm cell after 1mg/ml of ascorbic acid treatment, where panel (g-i) shows the focused, defocused, and reconstructed phase map image. Figure 3D represents the sperm cell after treating with 8mg/ml of ascorbic acid, where (j-l) represent focused, defocused, and reconstructed phase map.

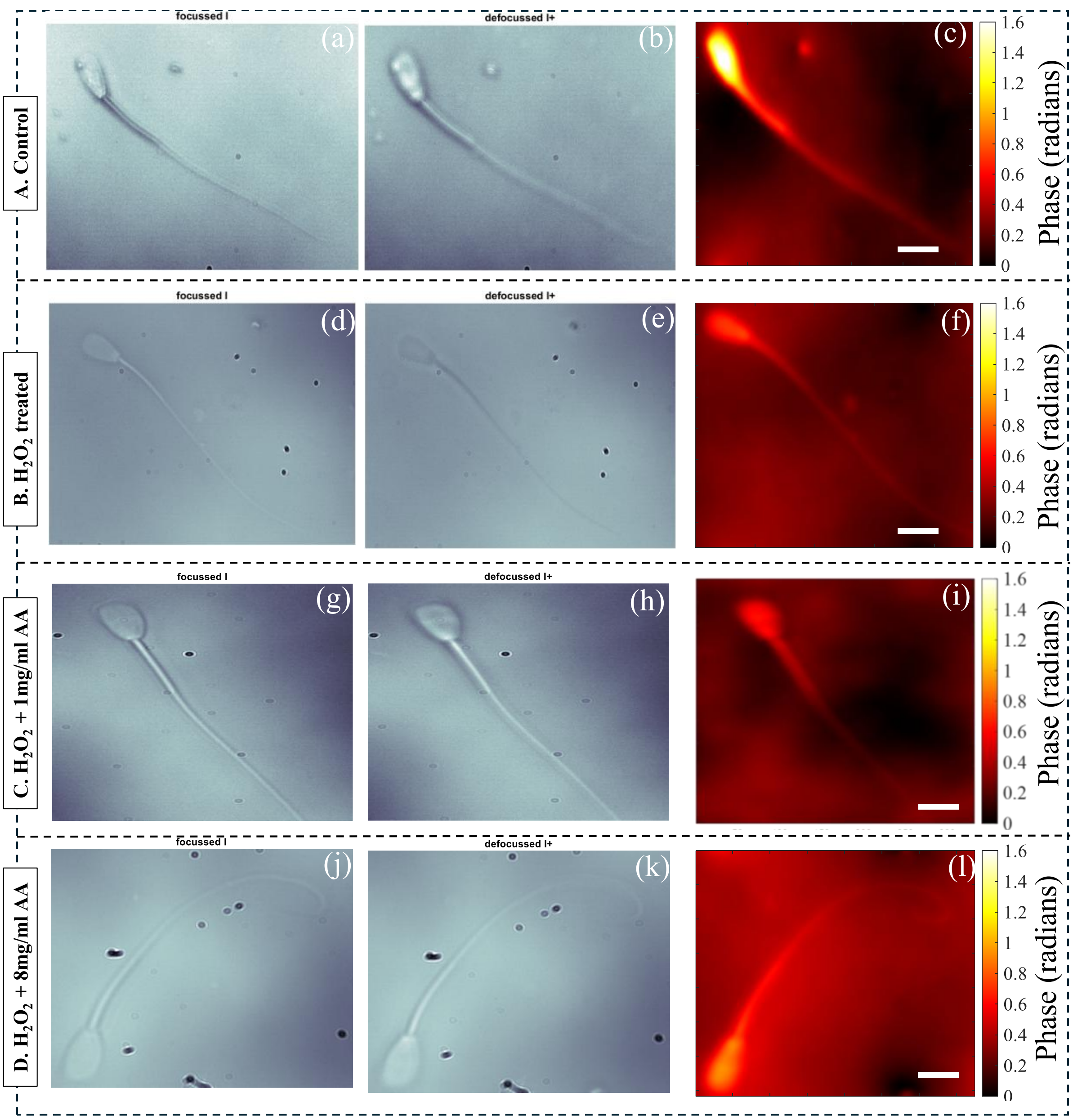

**Figure 3:** Representative quantitative phase microscopy (QPM) images of single bovine spermatozoa under three experimental conditions: (A) control, (B) oxidative stress induced by $H_2O_2$, and (C) antioxidant treatment with 1 mg/ml of ascorbic acid; (D) antioxidant treatment with 8mg/ml of ascorbic acid. For each condition, subfigure (a) shows the in-focus bright-field image, (b) the defocused image acquired at a 2 µm offset, and (c) the reconstructed

quantitative phase map obtained using the Transport of Intensity Equation (TIE). The color-coded phase distribution represents quantitative phase delays corresponding to intracellular refractive index and thickness. A reduction in phase amplitude is observed under oxidative stress, while partial recovery is evident with 8mg/ml ascorbic acid supplementation compared with 1mg/ml. Scale bar is in radians. A scale bar corresponds to 10µm. (AA- ascorbic acid)

Figure 4(a) presents the box plot distribution of optical thickness values across the four experimental groups. Exposure to $H_2O_2$ resulted in a statistically significant reduction in optical thickness compared to the control group ($p < 0.05$), consistent with oxidative stress-induced alterations in intracellular macromolecular density and membrane integrity within the sperm head. The reconstructed phase maps further confirm this finding, demonstrating a measurable decrease in phase values across the sperm head following $H_2O_2$ exposure.

Treatment with 1 mg/ml ascorbic acid did not result in a statistically significant recovery of optical thickness compared to the $H_2O_2$-only group, with values remaining comparable to those in the oxidatively stressed condition, suggesting that this concentration may provide insufficient antioxidant capacity under the acute oxidative conditions employed. In contrast, co-treatment with 8 mg/ml ascorbic acid resulted in a statistically significant partial increase in optical thickness compared with the $H_2O_2$-only group, with values trending toward, but not fully reaching, control levels, indicating partial preservation of intracellular structural integrity at the higher antioxidant concentration.

The optical thickness measurement in QPM reflects the local concentration of organic macromolecules integrated along the optical path through the cell. Within the sperm head, the nucleus is the region of high optical thickness due to its dense chromatin, distinguishing it from the comparatively thin acrosomal region [30,32]. The reconstructed phase maps confirmed the highest phase values in the nuclear region, consistent with its known structural composition. Quantification of optical thickness in the nuclear region, therefore, represents a sensitive and objective biophysical indicator of oxidative stress-induced structural changes and antioxidant-mediated partial recovery, as illustrated in Figure 4(a).

### 3.2 Morphological and textural evaluation of antioxidant recovery in oxidized bovine spermatozoa

Morphological analysis of QPM-derived three-dimensional parameters revealed a consistent pattern of statistically significant structural changes across all five other parameters evaluated: volume, dry mass, surface area, sphericity, and surface area-to-volume ratio (Figure 4b–f). In each case, the $H_2O_2$-only group demonstrated statistically significant reductions compared to the control, collectively consistent with oxidative stress-induced loss of intracellular biomass, membrane shrinkage, and morphological deformation of the sperm head. Treatment with 1 mg/ml ascorbic acid did not result in a statistically significant recovery across any of these parameters, with values remaining comparable to those of the oxidatively stressed group, suggesting that this concentration provides insufficient antioxidant capacity to counteract the degree of acute structural damage imposed by the experimental model. In contrast, co-treatment with 8 mg/ml ascorbic acid resulted in partial but statistically significant recovery across all five parameters compared to the $H_2O_2$-only group, though full restoration to control levels was not achieved for any parameter. Especially, the concurrent reduction in sphericity and increase in surface area-to-volume ratio in the $H_2O_2$-only group together indicate oxidative stress-induced structural deformation of the sperm head, consistent with morphological changes previously reported in oxidatively stressed spermatozoa. The partial attenuation of these changes at 8 mg/ml ascorbic acid further supports a dose-dependent structural protective effect, whereas the absence of recovery at 1 mg/ml suggests a concentration threshold below which ascorbic acid supplementation is insufficient to counteract the degree of acute structural damage imposed by the experimental model.

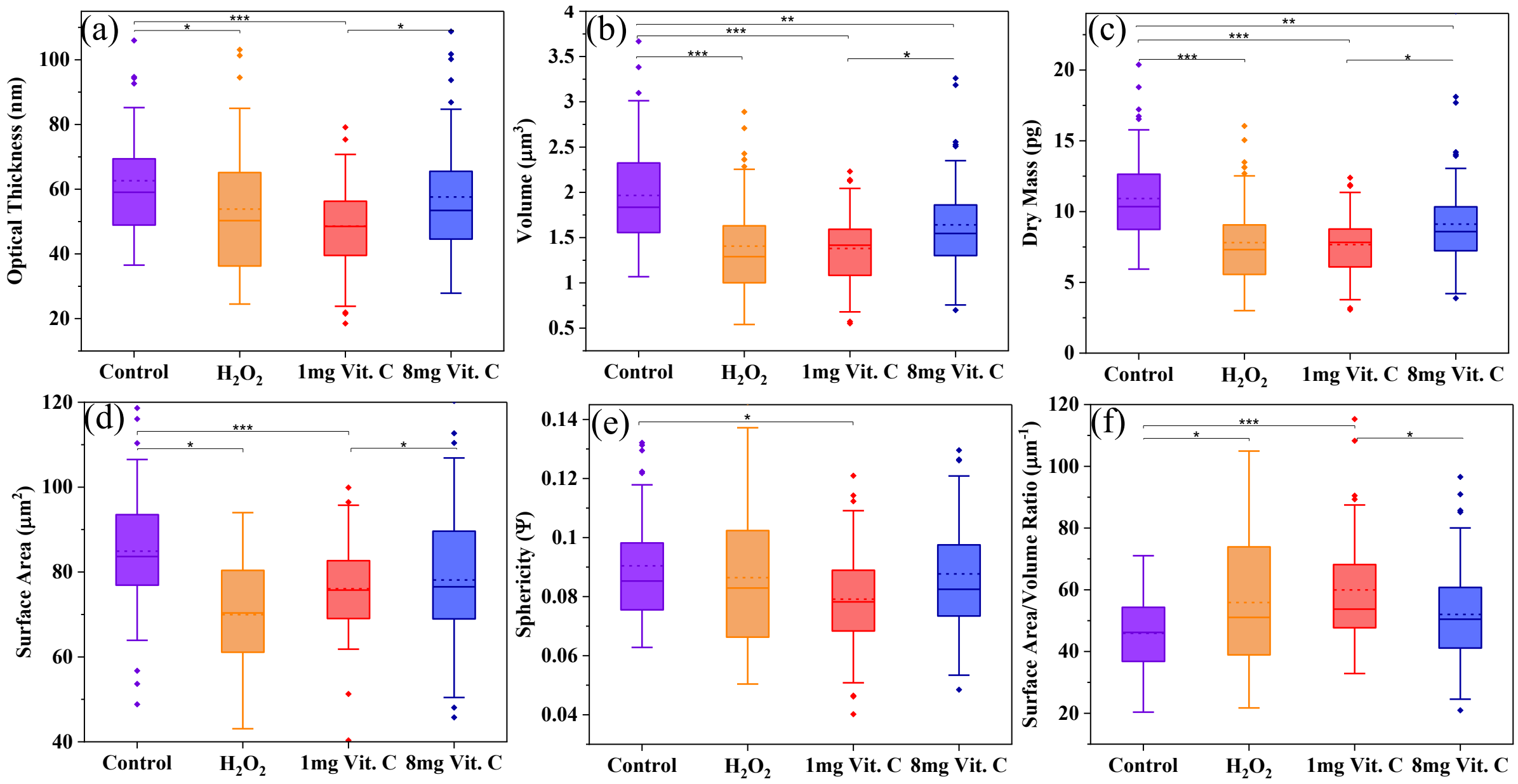

**Figure 4**: Quantitative evaluation of sperm head biophysical parameters under oxidative stress and ascorbic acid supplementation. Box plots illustrate group differences in (a) optical thickness, (b) optical volume, (c) dry mass, (d) surface area, (e) sphericity, and (f) surface area-to-volume ratio. Experimental groups are color-coded: Control (purple), $H_2O_2$-induced oxidative stress (orange), $H_2O_2$ + Ascorbic Acid 1 mg/ml (red), and $H_2O_2$ + Ascorbic Acid 8 mg/ml (blue); n = 60 cells per group. The central horizontal line represents the median, the dashed line indicates the mean, and the box boundaries define the interquartile range (25th–75th percentiles). Whiskers extend to 1.5 times the interquartile range, with individual points beyond this range shown as outliers. Statistical significance was assessed using a one-way ANOVA followed by Tukey's post hoc multiple-comparison test. Bracketed lines indicate specific statistically significant pairwise comparisons (* $p \leq 0.05$, ** $p \leq 0.01$, *** $p < 0.001$). Group pairs that did not reach statistical significance are not annotated.

Texture parameters extracted from optical thickness maps of segmented sperm head regions of interest provided complementary biophysical characterization of intracellular structural organization across experimental groups; mean values with standard deviations are presented in Table 2, and corresponding box plots are shown in Supplementary File 2. The mean optical thickness and standard deviation decreased in the $H_2O_2$-only group compared with the control, consistent with a reduction in the average intracellular structural heterogeneity across the sperm head, reflecting oxidative stress-induced loss of intracellular macromolecular density. Co-treatment with 1 mg/ml ascorbic acid did not result in statistically significant recovery toward control values, whereas the 8 mg/ml group demonstrated a partial increase in mean optical thickness and standard deviation, suggesting a degree of intracellular content preservation at the higher antioxidant concentration.

Skewness values shifted toward greater negative asymmetry in the $H_2O_2$-only group compared to the control, consistent with redistribution of intracellular phase values toward lower optical thickness regions, reflecting oxidative stress-induced loss of intracellular macromolecular content. Treating with 1 mg/ml ascorbic acid did not restore skewness, with values remaining statistically comparable to those in the $H_2O_2$-only group. In contrast, co-treatment with 8 mg/ml ascorbic acid resulted in a statistically significant recovery of skewness to levels comparable to those of the control group, suggesting effective preservation of intracellular phase distribution regularity at the higher antioxidant concentration.

Kurtosis values did not differ significantly between the control and $H_2O_2$-only groups, suggesting that acute oxidative stress under the conditions employed did not produce a measurable change in the peak of the intracellular phase distribution. However, treatment with 1 mg/ml ascorbic acid resulted in a statistically significant increase in kurtosis compared with all other groups. This unexpected finding at the lower ascorbic acid concentration warrants careful interpretation and may reflect an indirect structural effect at this dose rather than a protective response. Co-treatment with 8 mg/ml ascorbic acid did not produce values significantly different from the control, consistent with structural preservation at the higher concentration.

Textural entropy showed stable values across the control, $H_2O_2$-only, and 8 mg/ml ascorbic acid groups, with no statistically significant differences among them. A slight but statistically significant reduction was observed in the 1 mg/ml ascorbic acid group compared to the control, though the biological significance of this isolated finding requires further investigation. The overall stability of entropy across the remaining groups suggests that while acute $H_2O_2$-induced oxidative stress significantly reduced the magnitude of optical thickness and dry mass, it did not substantially alter the spatial randomness of the optical thickness distribution within the sperm head. This parameter-specific finding indicates that the oxidative stress induced an overall reduction in intracellular optical thickness magnitude without causing statistically detectable spatial reorganization of the optical thickness distribution under the acute conditions employed.

Textural energy demonstrated the most pronounced and consistent response across experimental groups. The $H_2O_2$-only group exhibited a statistically significant increase in energy compared with all other groups, consistent with a shift toward greater intracellular phase uniformity and loss of distinct subcellular boundaries, suggesting cytoplasmic homogenization under acute oxidative conditions. In particular, both ascorbic acid treatment groups (1 mg/ml and 8 mg/ml) demonstrated statistically significant attenuation of the $H_2O_2$-induced energy increase, with values numerically intermediate between the control and $H_2O_2$-only groups but not statistically significantly different from the control, indicating partial attenuation of the oxidative stress-induced shift toward intracellular phase uniformity at both ascorbic acid concentrations used.

| **Parameters** | **Control** | **$H_2O_2$ treated** | **$H_2O_2$ + 1mg of AA** | **$H_2O_2$ + 8mg of AA** |
|---|---|---|---|---|

| Optical thickness (nm) | 62.61 ± 19.06$^{a}$ | 53.85 ± 20.84$^{bc}$ | 48.58 ± 12.90$^{c}$ | 57.58 ± 19.63$^{ab}$ |
|---|---|---|---|---|
| Volume ($\mu m^3$) | 1.96 ± 0.58$^{a}$ | 1.40 ± 0.53$^{bc}$ | 1.37 ± 0.38$^{c}$ | 1.64 ± 0.61$^{b}$ |
| Dry mass (pg) | 10.91 ± 3.25$^{a}$ | 7.80 ± 2.96$^{bc}$ | 7.66 ± 2.16$^{c}$ | 9.11 ± 3.43$^{b}$ |
| Surface area ($\mu m^2$) | 84.91 ± 14.83$^{a}$ | 69.95 ± 13.38$^{bc}$ | 76.05 ± 10.11$^{c}$ | 78.12 ± 16.79$^{ab}$ |
| Sphericity (ψ) | 0.090 ± 0.02$^{a}$ | 0.088 ± 0.02$^{ab}$ | 0.079 ± 0.01$^{b}$ | 0.087 ± 0.02$^{ab}$ |
| Surface area/Volume ratio (S/V) ($\mu m^{-1}$) | 45.89 ± 11.38$^{a}$ | 55.89 ± 20.30$^{bc}$ | 59.94 ± 20.94$^{c}$ | 52.01 ± 16.63$^{ab}$ |
| Mean | 47.0 ± 14.69$^{a}$ | 41.13 ± 15.96$^{ab}$ | 36.60 ± 10.20$^{bc}$ | 42.80 ± 14.96$^{a}$ |
| Standard deviation | 8.77 ± 2.52$^{a}$ | 7.15 ± 2.75$^{bc}$ | 6.38 ± 1.84$^{c}$ | 8.11 ± 2.71$^{ab}$ |
| Skewness | -0.048 ± 0.14$^{ab}$ | -0.108 ± 0.15$^{b}$ | -0.106 ± 0.16$^{b}$ | -0.026 ± 0.13$^{a}$ |
| Kurtosis | 1.85 ± 0.11$^{b}$ | 1.87 ± 0.11$^{b}$ | 1.98 ± 0.16$^{a}$ | 1.86 ± 0.13$^{b}$ |
| Entropy | 3.83 ± 0.030$^{a}$ | 3.82 ± 0.03$^{a}$ | 3.80 ± 0.04$^{b}$ | 3.83 ± 0.03$^{a}$ |
| Energy | $3.46 \times 10^{-4} \pm 6.46 \times 10^{-5}$ $^{b}$ | $4.23 \times 10^{-4} \pm 1.02 \times 10^{-4}$ $^{a}$ | $3.80 \times 10^{-4} \pm 5.91 \times 10^{-5}$ $^{b}$ | $3.85 \times 10^{-4} \pm 9.92 \times 10^{-5}$ $^{b}$ |

**Table 2**: Biophysical and texture parameters of sperm cell heads (Mean ± Standard deviation) under normal (control) conditions, after external oxidative stress induction, following antioxidant recovery with 1mg/ml of ascorbic acid, and with 8mg/ml of ascorbic acid. Values are presented as Mean ± SD. Different superscript letters (a, b, c) within the same row indicate statistically significant differences between experimental groups ($p < 0.05$), as determined by one-way ANOVA followed by Tukey's post-hoc test. (AA-ascorbic acid)

## 4. Discussion

1. **QPM-based derived parameters as an indicator of oxidative stress**: The primary objective of this study was to investigate whether ascorbic acid can attenuate acute oxidative stress-induced structural changes in bovine spermatozoa, as detected by QPM-derived biophysical parameters. The high-resolution, label-free phase maps generated by TIE-based

QPM revealed structural differences across experimental groups that are consistent with $H_2O_2$-induced oxidative damage and suggestive of partial dose-dependent structural preservation following ascorbic acid supplementation. These findings indicate that QPM-derived parameters, including optical thickness, dry mass, sphericity, and intracellular texture indices, are sensitive to oxidative stress-induced structural changes and may serve as promising biophysical indicators of the degree of structural attenuation conferred by antioxidant treatment.

**2. Dose-dependent structure recovery by ascorbic acid**: Establishing the optimal ascorbic acid dosage was crucial to effectively counteract rigorous $H_2O_2$ stress. This high-stress scenario was intentionally designed to produce rapid, clear 3D structural collapse, ideal for high-resolution phase imaging. To determine the most effective dose for recovery, we tested rescue formulations containing 1 mg/ml and 8 mg/ml of ascorbic acid. The results were decisive: the 1 mg/ml formulation provided minimal structural protection under these extreme conditions and showed inconsistent structural recovery across QPM-derived biophysical and texture parameters. For several parameters, values in this group remained statistically comparable to the $H_2O_2$ only oxidative stress group rather than showing meaningful recovery toward control levels, suggesting that this concentration may provide insufficient antioxidant capacity to counteract the degree of acute structural damage imposed by the experimental model. These findings highlight that the protective effect of ascorbic acid is both dose-dependent and parameter-specific, and caution against generalizing any protective effect without specifying the effective concentration range.

In contrast, the 8 mg/ml ascorbic acid formulation demonstrated significant structural protection as evidenced by statistically significant improvements in selected QPM-derived morphological and texture parameters compared to the oxidative stress group. However, complete restoration of all parameter values to control levels was not observed. This partial, dose-dependent structural attenuation observed in our QPM-derived biophysical parameters is highly consistent with the antioxidant literature [33]. While significant improvements following ascorbic acid treatment are frequently reported, complete restoration of all parameters to baseline levels is rarely demonstrated under conditions of acute and severe oxidative challenge [20,34]**.** Previous studies employing $H_2O_2$ and heat-induced oxidative stress models have demonstrated that ascorbic acid reduces the oxidation-reduction potential in a dose-dependent manner; however, significant functional reductions often persist, indicating that protective effects are partial rather than absolute [34–36]. Under severe oxidative stress, when high ROS levels overwhelm cellular defenses, complete recovery of all structural and functional parameters cannot be expected from a single antioxidant intervention. Ultimately, this underscores the necessity of a high-concentration therapeutic threshold (such as our 8 mg/ml formulation) to effectively combat severe oxidative stress and preserve cellular topography.

The observed decreases in optical thickness and dry mass in the $H_2O_2$-only group are consistent with structural changes associated with membrane disruption and intracellular protein loss, findings suggestive of lipid peroxidation-mediated damage, given the known vulnerability of sperm plasma membrane polyunsaturated fatty acids to ROS-induced peroxidation. Since optical path difference reflects the local concentration of organic macromolecules within the cell, the relative maintenance of phase delay values in the 8 mg/ml co-treatment group compared to the $H_2O_2$-only group suggests that higher-dose ascorbic acid may partially preserve membrane barrier function and intracellular protein content under acute oxidative conditions, consistent with its known role as an electron donor capable of scavenging extracellular and intracellular ROS when present simultaneously with the oxidative stress

**3. Limitations:** The classical biochemical markers of oxidative stress were not evaluated in the present study; these observations should be interpreted as structural and biophysical indicators rather than biochemical or functional confirmation of antioxidant efficacy. Additionally, the absence of an ascorbic acid-only control group limits the interpretation of the independent structural effect of ascorbic acid in the absence of oxidative stress. Future studies should include this additional control condition to fully characterize the concentration-dependent effects of ascorbic acid on sperm structural parameters independent of oxidative challenge. Furthermore, while QPM-derived textural parameters are highly sensitive indicators of cellular stress, they represent mathematical descriptors of physical disorder rather than direct molecular readouts; therefore, concluding specific biological mechanisms from these metrics should be interpreted with caution. Another limitation of this study is that semen was obtained from a single bull, and the n = 60 cells per group, therefore, represent technical rather than biological replication. While this controlled design was intentional, it limits the validity of the findings. Future studies incorporating semen from multiple bulls across independent experimental sessions are recommended to provide true biological replication and further validate the QPM-based assessment framework reported here.

**4. Texture parameters as an indicator of oxidative stress:** Among the texture parameters evaluated, energy emerged as the most consistently responsive indicator of oxidative stress-induced structural change, suggesting its potential utility as a discriminative label-free biomarker in sperm quality assessment. The paradoxical increase in kurtosis observed specifically in the 1 mg/ml ascorbic acid group warrants careful interpretation and may reflect indirect structural effects at sub-optimal antioxidant concentrations, representing an important direction for future investigation. The overall stability of entropy across experimental groups suggests that this parameter reflects spatial reorganization of the optical thickness distribution rather than a reduction in magnitude and may therefore be less sensitive to acute oxidative damage of the type induced in the present model.

## 5. Conclusion

In conclusion, this study demonstrates that QPM-derived biophysical and texture parameters are sensitive to $H_2O_2$-induced structural changes in bovine spermatozoa and can detect dose-dependent differences in structural preservation following ascorbic acid co-treatment. The 8 mg/ml ascorbic acid formulation conferred partial, but statistically significant, attenuation of oxidative stress-induced structural degradation across selected QPM-derived parameters, while the 1 mg/ml formulation demonstrated limited and inconsistent protection, underscoring the concentration-dependent nature of ascorbic acid's structural protective capacity. These findings suggest that QPM may serve as a promising, label-free tool for the structural assessment of spermatozoa under oxidative challenge, providing biophysical information not accessible through conventional sperm evaluation techniques. Future studies incorporating semen from multiple bulls, classical oxidative stress markers, and direct functional validation, including motility analysis, DNA integrity assessment, and in vitro fertilization trials, are necessary to determine whether the structural preservation demonstrated here translates to improved outcomes in ART protocols.

## Acknowledgement

H Kaur would like to acknowledge the Ministry of Human Resource Development (MHRD) for financial support during the research work through the fellowship and the Ministry of Science and Technology, Department of Biotechnology, Government of India, Project No. BT/PR50402/AAQ/1/986/2023.

## Declaration of interests

The authors declare no competing interests.